\documentclass{an}
\usepackage{graphicx}
\usepackage{times}
\Volume{01}              % issue in current volume (01...06)
\Year{2003}              % format: {YYYY}
\Month{10}               % format: {M} or {MM} (number), month of
\Pagespan{1}{2}		 % format: {start page}{last page}

\def\sauron{{\tt SAURON}}
\def\lesssim{\mathrel{\hbox{\rlap{\hbox{\lower4pt\hbox{$\sim$}}}\hbox{$<$}}}}
\def\farcs{\hbox{$.\!\!^{\prime\prime}$}}
\def\oiii{[O\,{\sc iii}]}
\def\halpha{\hbox{H$\alpha$}}
\def\hbeta{\hbox{H$\beta$}}

\def\aj{AJ}             	% Astronomical Journal
\def\apj{ApJ}           	% Astrophysical Journal
\def\apjl{ApJ}          	% Astrophysical Journal, Letters
\def\apjs{ApJS}         	% Astrophysical Journal, Supplement
\def\aap{A\&A}          	% Astronomy and Astrophysics
\def\mnras{MNRAS}       	% Monthly Notices of the RAS
\begin{document}
\lhead[\thepage]{Falc\'on-Barroso et al.: A \sauron\ look at galaxy bulges}
\rhead[Astron. Nachr./AN~{\bf XXX} (200X) X]{\thepage}
\headnote{Astron. Nachr./AN {\bf 32X} (200X) X, XXX--XXX}

\title{A \sauron\ look at galaxy bulges}

\author{J.~Falc\'on-Barroso\inst{1}, 
R.~Bacon\inst{2}, M.~Bureau\inst{3},
M.~Cappellari\inst{1}, R.~L.~Davies\inst{4},
E.~Emsellem\inst{2}, D.~Krajnovi\'c\inst{1},
H.~Kuntschner\inst{5}, R.~McDermid\inst{1},
R.~F.~Peletier\inst{6} \and P.~T.~de~Zeeuw\inst{1}} 

\institute{Sterrewacht Leiden, Niels Bohrweg~2, 2333~CA, Leiden, The Netherlands
\and Centre de Recherche Astronomique de Lyon, 9~Avenue Charles--Andr\'e, 69230 Saint-Genis-Laval, France
\and Columbia Astrophysics Laboratory, 550 West 120$^{th}$ Street, 1027 Pupin Hall, MC~5247, New York, NY~10027, U.S.A.
\and Denys Wilkinson Building, University of Oxford, Keble Road, Oxford, United Kingdom
\and European Southern Observatory, Karl Schwarzschild Strasse~2, D-85748 Garching, Germany
\and Kapteyn Astronomical Institute, University of Groningen, Postbus 800, 9700 AV  Groningen, The Netherlands}

\date{Received {\it date will be inserted by the editor}; accepted {\it date will be inserted by the editor}} 

\abstract{Kinematic and population studies show that bulges are generally 
rotationally flattened systems similar to low-luminosity ellipticals.
However, observations with state-of-the-art integral field spectrographs,
such as \sauron, indicate that the situation is much more complex, and allow us
to investigate phenomena such as triaxiality, kinematic decoupling and population 
substructure,  and to study their connection to current formation and evolution 
scenarios for bulges of early-type galaxies. We present the examples of two S0 
bulges from galaxies in our sample of nearby galaxies: one that shows all the 
properties expected from classical bulges (NGC\,5866), and another case that 
presents kinematic features appropriate for barred disk galaxies (NGC\,7332).
\keywords{galaxies: evolution -- galaxies: formation --
	  galaxies: elliptical and lenticular, cD --  
	  galaxies: kinematics and dynamics --
	  galaxies: individual (NGC\,5866, NGC\,7332)}}

\correspondence{jfalcon@strw.leidenuniv.nl}

\maketitle

\section{Introduction}
Lying at the centre of galaxies, bulges are a keystone in our
understanding of galaxy formation and evolution. When studying their
formation, one is often drawn to examine the order of events during galaxy
assembly. In a simplistic approach, either the bulge formed before the
disk, or the bulge formed from disk material.

Classical bulges are traditionally understood as spheroids with an 
$\mathrm{r}^{1/4}$ surface brightness profile, in most ways identical to
elliptical galaxies of the same luminosity. For example, bulges behave like
ellipticals in the widely known scaling relations  (Faber \& Jackson 1976;
Terlevich et al. 1981;  Djorgovski \& Davis 1987; Dressler et al. 1987). 
Kinematically,  bulges, like low-luminosity ellipticals, are found to be 
isotropic oblate rotators  (Kormendy \& Illingworth 1982). Such analogies  lie at
the heart of merger-driven bulge formation  models, in which the bulge formed 
before the disk.

Dynamicists, however, have shown in their numerical simulations that disk
instabilities can pump disk material above the plane,  thus generating
central structures that also {\it bulge} over the  thin disk (e.g. Hasan,
Pfenniger \& Norman 1993). These simulations raised the  possibility of an
alternative formation scenario in which bulges are formed  via secular
evolution processes after the disk  (e.g. Pfenniger 1993).  The presence
of bars inside peanut-shaped bulges has been confirmed photometrically 
(L{\" u}tticke, Dettmar \& Polhen 2000)  and spectroscopically  
(Kuijken \& Merrifield 1995; Bureau \& Freeman 1999; Merrifield \& Kuijken 1999). 
Furthermore, some bulges  share strong similarities with disks, on the basis of 
their low velocity dispersions. The disky nature of bulges is usually discussed 
in relation to late-type, low-mass bulges (Kormendy 1993, Carollo 1999).

Do bulges come in two flavours, early-types forming in mergers and
late-types through disk instabilities? To answer this question, it is
vital to increase our knowledge on the structural and dynamical properties
that link galaxy bulges to either slowly rotating spheroids akin to
elliptical galaxies, or to rapidly-rotating flattened systems more nearly
resembling the products of internal  disk transformations.

Advances and discoveries in astrophysics often come together with
instrument developments. Photometrically, the detailed pictures HST
provided during the last decade, has revealed a large variety  of
features (e.g. inner disks or bars) at the centre of galaxies  (Lauer et
al. 1995; Carollo 1999; Balcells et al. 2003). Spectroscopically, 
however, it is the ground-based integral field units (IFUs) that 
can map the full 2D stellar and gas kinematic signatures of those 
structures. This allows us to impose tighter constraints on the
processes involved in the formation and evolution of galaxies. 

The \sauron\ survey is a scientific project aiming to understand
the formation  and evolution of elliptical and lenticular galaxies and of
spiral bulges from 2D observations of a representative sample of E/S0 
galaxies and Sa bulges, using a custom-built panoramic integral field
spectrograph placed at the 4.2m WHT on La Palma (Spain). Our early 
results (Bacon et al. 2001; de Zeeuw et al. 2002; Emsellem et al. 2003)
reveal a variety of structures much richer than usually recognized in
E/S0s galaxies. 

We are carrying out a number of collaborative projects to investigate 
also bulges of later type (e.g. Sbc). These are taken from the samples 
of Balcells \& Peletier (1994) and Carollo et al. (1997), which have 
extensive HST imaging available and in some cases also long-slit 
spectroscopy (e.g. Carollo et al. 2002; Balcells et al. 2003; 
Falc\'on-Barroso et al. 2003a).

Here we present two examples of galaxy bulges observed with \sauron.  
One case (NGC\,5866) shows all the signatures expected on a classical 
bulge (see Sec. 1), whereas the second one (NGC\,7332) presents properties 
of barred disks galaxies.

\section{NGC\,5866: a classical bulge}
\begin{figure*}
\resizebox{\hsize}{!}{\includegraphics[angle=0]{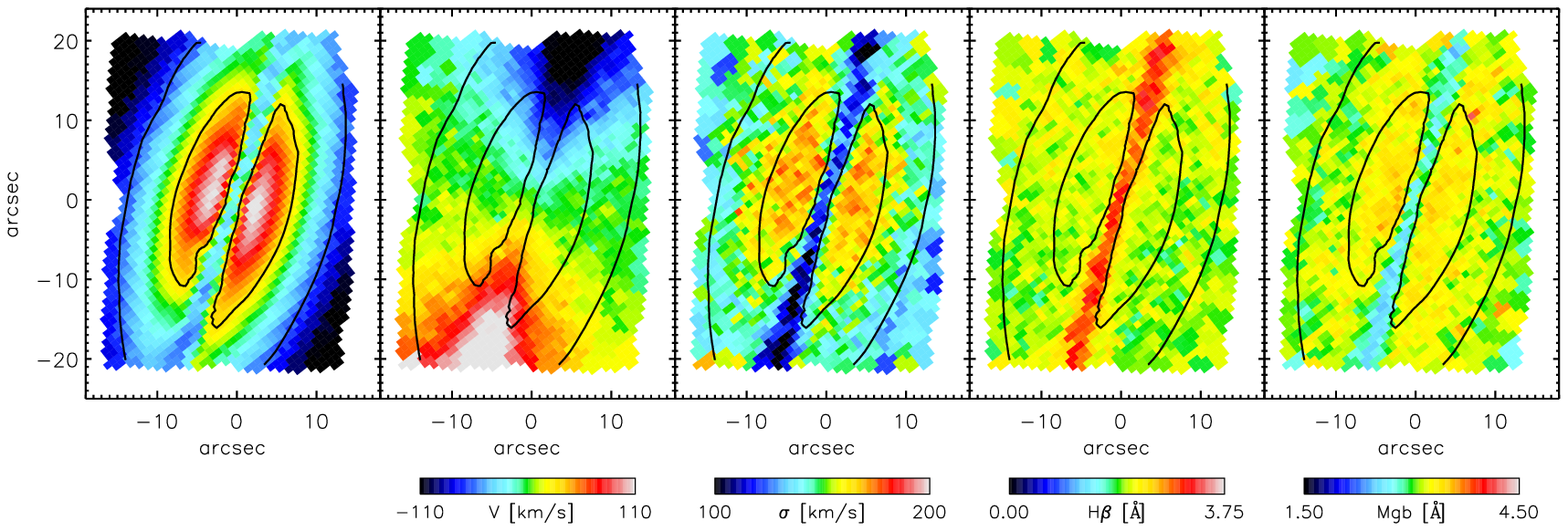}}
\caption{\sauron\ maps of the S0/a galaxy NGC\,5866. From left to right: 
integrated intensity, radial velocity of the stars, stellar velocity
dispersion, \hbeta\ and Mg$b$ line-strength indices (both on  the Lick
system). The absorption line maps have been determined after separating
absorption and emission (Emsellem et al. 2003; Falc\'on-Barroso et al.
2003b).}
\label{Fig:N5866}
\bigskip
\bigskip
\end{figure*}

NGC\,5866 is an edge-on S0 galaxy that has been studied in some of the
most recent surveys of galaxies (Filho, Barthel \& Ho 2002; McMahon et al.
2002; Terlevich \& Forbes 2002; Jarrett et al. 2003). NGC\,5866 shows a
large bulge and an edge-on disk with a prominent dust lane
(Fig.~\ref{Fig:N5866}). The multi-slit analysis of its stellar kinematics
shows a smoothly rising, featureless rotation curve (Fisher 1997), and the
ionized-gas behaves like its stellar counterpart (Fisher 1997).   

The power of the \sauron\ observations to obtain simultaneously 2D spatial and
spectroscopic information allows us to show that the bulge and disk in NGC\,5866  are
separated both photometrically and kinematically. As expected in classical 
bulges, the disk is a flattened, rapidly rotating structure, while the bulge is a 
hotter, rounder and slower rotating component. The bulge does not rotate 
cylindrically, unlike the case of NGC\,7332 (see below), but as normal spheroids. 
The large increase in the \hbeta\ line index in the disk, compared to the bulge 
(see Fig.~\ref{Fig:N5866}), means that the most recent stars formed
in the disk. This age difference is also apparent from the decrease of the Mg$b$ index.
The combination of dynamical and stellar population models in this galaxy, but
also for the others in the sample, will prove extremely important to put
strong constraints on formation scenarios of bulges.

\begin{figure}
\resizebox{\hsize}{!}{\includegraphics[angle=90]{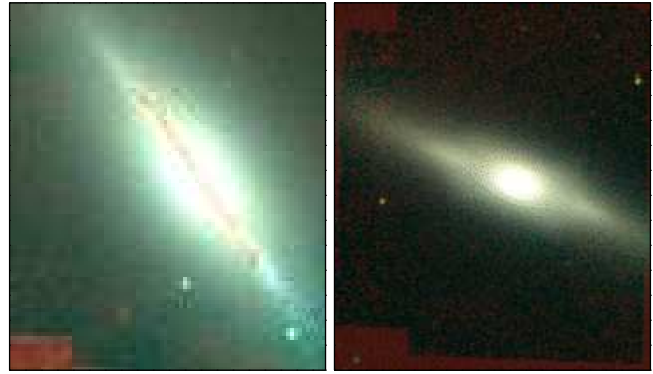}}
\caption{$B$-$R$-$K$ composite images (Peletier \& Balcells 1997) of 
NGC\,5866 (left panel) and NGC\,7332 (right panel).}
\label{Fig:sample}
\end{figure}

\begin{figure*}
\resizebox{\hsize}{!}{\includegraphics[angle=0]{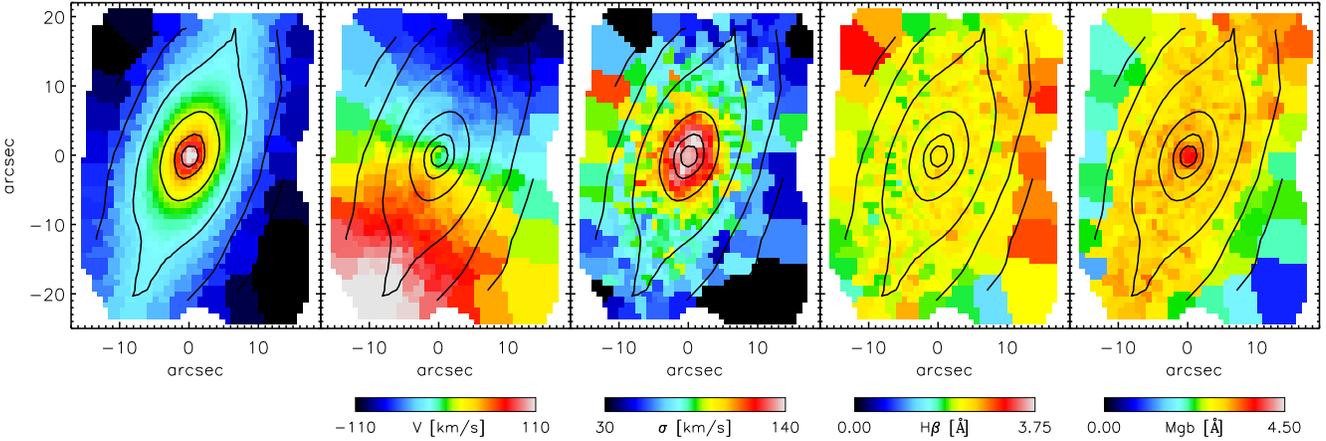}}
\caption{\sauron\ maps of the S0 galaxy NGC\,7332. From left to right: 
integrated intensity, radial velocity of the stars, stellar velocity
dispersion, \hbeta\ and Mg$b$ line-strength indices (both on the Lick
system). The absorption line maps have been determined after separating
absorption and emission (Emsellem et al. 2003; Falc\'on-Barroso et al.
2003b).The \sauron\ spectra have been  spatially binned to a minimum $S/N$
of $60$ by means of the Voronoi 2D binning  algorithm of Cappellari \&
Copin (2003).}
\label{Fig:N7332}
\bigskip
\bigskip
\end{figure*}

\section{NGC\,7332: a bulge made up from a bar}
NGC\,7332 is an ordinary looking edge-on S0 galaxy (Fig.~\ref{Fig:sample})
that has been extensively studied in the past. Photometric analysis
reveals a boxy bulge, a central disk and  evidence for the presence of a
`weak` bar (Seifert \& Scorza 1996; L{\" u}tticke et al. 2000). However,
the galaxy is mainly known for a  bright counter-rotating and a faint
co-rotating \oiii\ gas component with  respect to the stars  (Bertola,
Buson \& Zeilinger 1992; Fisher, Illingworth \& Franx 1994). Those gas
structures were confirmed by Plana \& Boulesteix (1996), who mapped the
\halpha\ emission via   Fabry-Perot observations. NGC\,7332 colours are
somewhat bluer than those of elliptical galaxies of the same luminosity.
Spectral analysis of the central regions reveals a luminosity-weighted age
of about $6$~Gyr (Vazdekis \& Arimoto 1999).  

The \sauron\ stellar kinematics (see Fig.~\ref{Fig:N7332}) displays a
rather smooth velocity field with  rotation along  the major-axis and a
weak dependence of rotation on galactic height. The stellar kinematics
also shows, for the first time, a decoupled component in the centre, 
misaligned with respect to the galaxy's kinematic major-axis, which may 
be related to a dip in the velocity dispersion map also in the centre 
($\rm{r}\lesssim2\farcs5$). As shown in Falc\'on-Barroso et al. (2003b), 
the gas exhibits very complex morphology and kinematics. This is found 
especially in \oiii\, which is again mainly counter-rotating with respect 
to the  stars. The analysis of the absorption line-strengths reveal that
NGC\,7332's stellar populations are generally young ($5\pm2$~Gyr), not only in the
disk but also in the bulge, in agreement with previous studies (Balcells
\& Peletier 1994; Vazdekis et al. 1996; Terlevich \& Forbes 2002). The
metal absorption lines (e.g. Mg$b$) show  an increase in the centre,
contrasting with the rather homogeneous \hbeta\ index. As  emphasized by
Falc\'on-Barroso et al. (2003b), the unique data set provided by \sauron
provides evidence for a formation scenario where both bar-driven processes 
and interactions play a significant role.

\section{Conclusions}
Bulges are not just simple scaled-down versions of elliptical galaxies. Instead, 
they often present properties closely related to bars or disks, that argue 
for different formation mechanisms.
Integral field spectrographs have the ability to map the 2D behaviour of 
galaxies and allow us to characterize the full dynamical state of 
the bulges. Features like cylindrical rotation, triaxiality, kinematic 
decoupling, bars, as well as the amount of rotational support as a function 
of distance above the plane can be easily studied. Furthermore, the 
line-strength maps contain essential information on the spatial distribution 
of the stellar populations, allowing us to determine global population gradients
in the radial and vertical directions. All these capabilities, combined with 
dynamical and stellar population models, will provide unprecedented observational 
information that can be used to constrain the formation and evolution processes in 
bulges of galaxies.

\acknowledgements
This work is based on observations obtained at the WHT on the island of La
Palma, operated by the Isaac Newton Group at the Observatorio del Roque
de los Muchachos of the Instituto de Astrof\'\i sica de Canarias.


\begin{thebibliography}{}
\bibitem{}
{Bacon}, R.,  {Copin}, Y.,  {Monnet}, G., et al.: 2001, \mnras~326, 23

\bibitem{}
{Balcells}, M., {Graham}, A.W., {Dom{\'{\i}}nguez-Palmero}, L., {Peletier}, R.F.: 2003, \apjl~582, L79

\bibitem{}
{Balcells}, M.,  {Peletier}, R.F.: 1994, \aj~107, 135

\bibitem{}
{Bertola}, F., {Buson}, L.M., {Zeilinger}, W.W.: 1992, \apjl~401, L79

\bibitem{}
{Bureau}, M.,  {Freeman}, K.C.: 1999, \aj~118, 126

\bibitem{}
{Cappellari}, M.,  {Copin}, Y.: 2003, \mnras~342, 345

\bibitem{}
{Carollo}, C.M.: 1999, \apj~523, 566


\bibitem{}
{Carollo}, C.M, {Stiavelli}, M., {de Zeeuw}, P.T., {Mack}, J.: 1997, AJ~114, 2366

\bibitem{}
{Carollo}, C.M, {Stiavelli}, M., {Seigar}, M., {de Zeeuw}, P.T., {Dejonghe}, H.: 2002, AJ~123, 159

\bibitem{}
{Djorgovski}, S., {Davis}, M.: 1987, \apj~313, 59

\bibitem{}
{Dressler}, A., {Lynden-Bell}, D., {Burstein}, D., et al.: 1987, \apj~313, 42

\bibitem{}
{de Zeeuw}, P.T.,  {Bureau}, M.,  {Emsellem}, E.,  et al.: 2002, \mnras~329, 513

\bibitem{}
{Emsellem}, E., {Cappellari}, M., {Peletier}, R.F., et al.: 2003, submitted to \mnras

\bibitem{}
{Faber}, S.M.,  {Jackson}, R.E.:  1976, \apj~204, 668

\bibitem{}
{Falc{\' o}n-Barroso}, J.,  {Balcells}, M.,  {Peletier}, R.F., {Vazdekis}, A.:  2003a, \aap~405, 455

\bibitem{}
{Falc{\' o}n-Barroso}, J., {Peletier}, R.F., {Emsellem}, E., et al.: 2003b, submitted to \mnras

\bibitem{}
{Filho}, M.E.,  {Barthel}, P.D.,    {Ho}, L.C.:  2002, \apjs~142, 223

\bibitem{}
{Fisher}, D.:  1997, \aj~113, 950

\bibitem{}
{Fisher}, D.,  {Illingworth}, G.,    {Franx}, M.:  1994, \aj~107, 160

\bibitem{}
{Hasan}, H.,  {Pfenniger}, D.,    {Norman}, C.:  1993, \apj~409, 91

\bibitem{}
{Jarrett}, T.H.,  {Chester}, T.,  {Cutri}, R.,  {Schneider}, S.E., {Huchra}, J.P.:  2003, \aj~125, 525

\bibitem{}
{Kormendy}, J.:  1993, in IAU Symp. 153: Galactic Bulges Vol.~153, 
{Kinematics of extragalactic bulges: evidence that some bulges are really disks}. p.~209

\bibitem{}
{Kormendy}, J.,  {Illingworth}, G.: 1982, \apj~256, 460

\bibitem{}
{Kuijken}, K.,  {Merrifield}, M.R.:  1995, \apjl, 443, L13

\bibitem{}
{Lauer}, T.R., {Ajhar}, E.A., {Byun}, Y.I., et al.: 1995, \aj~110, 2622

\bibitem{}
{L{\" u}tticke}, R.,  {Dettmar}, R.J.,    {Pohlen}, M.:  2000, \aap~362, 435

\bibitem{}
{McMahon}, R.G.,  {White}, R.L.,  {Helfand}, D.J.,    {Becker}, R.H.:  2002, \apjs~143, 1

\bibitem{}
{Merrifield}, M.R.,  {Kuijken} K.:  1999, \aap~345, L47

\bibitem{}
{Peletier}, R.F.,  {Balcells}, M.:  1997, New Astronomy~1, 349

\bibitem{}
{Pfenniger}, D.:  1993, in IAU Symp. 153: Galactic Bulges Vol.~153. p.~387

\bibitem{}
{Plana}, H.,  {Boulesteix}, J.:  1996, \aap~307, 391

\bibitem{}
{Seifert}, W.,  {Scorza}, C.: 1996, \aap~310, 75

\bibitem{}
{Terlevich}, A.I.,  {Forbes}, D.A.: 2002, \mnras~330, 547

\bibitem{}
{Terlevich}, R.,  {Davies}, R.L.,  {Faber}, S.M.,    {Burstein}, D.: 1981, \mnras~196, 381

\bibitem{}
{Vazdekis}, A.,  {Arimoto}, N.: 1999, \apj~525, 144

\bibitem{}
{Vazdekis}, A.,  {Casuso}, E.,  {Peletier}, R.F.,    {Beckman}, J.~E.: 1996, \apjs~106, 307

\end{thebibliography}
\end{document}